\documentclass[journal]{IEEEtran}
\pdfoutput=1
\usepackage[subpreambles=false]{standalone}
\usepackage{plasqn}
\usepackage{import}
\usepackage{eso-pic}

\IEEEoverridecommandlockouts
\graphicspath{{img/}{sections/img/}}
\begin{document}
	\title{On the Relationship Between Iterated Statistical Linearization and Quasi--Newton Methods}

	\author{
		\IEEEauthorblockN{Anton Kullberg, Martin A. Skoglund, Isaac Skog,~\IEEEmembership{Senior Member,~IEEE}, \\%
        and Gustaf Hendeby,~\IEEEmembership{Senior Member,~IEEE}.\\%
	\thanks{\noindent This work was partially supported by the Wallenberg AI,
		Autonomous Systems and Software Program (\textsc{WASP}) funded
		by the Knut and Alice Wallenberg Foundation.}
        \thanks{Anton Kullberg, Martin Skoglund, and Gustaf Hendeby are all with Link\"oping University, Sweden. Martin A. Skoglund is also affiliated with Eriksholm Research Centre, Snekkersten, Denmark. (email: \{anton.kullberg, martin.skoglund, gustaf.hendeby \}@liu.se). } 
        \thanks{Isaac Skog is with Uppsala University, Sweden. (email: isaac.skog@angstrom.uu.se).}
	}
  }

\maketitle

\begin{abstract}
This letter investigates relationships between iterated filtering algorithms based on statistical linearization, such as the iterated unscented Kalman filter (\abbrIUKF), and filtering algorithms based on quasi--Newton (\abbrQN) methods, such as the \abbrQN iterated extended Kalman filter (\abbrQNIEKF).
Firstly, it is shown that the \abbrIUKF and the iterated posterior linearization filter (\abbrIPLF) can be viewed as \abbrQN algorithms, by finding a Hessian correction in the \abbrQNIEKF such that the \abbrIPLF iterate updates are identical to that of the \abbrQNIEKF.
Secondly, it is shown that the \abbrIPLF/\abbrIUKF update can be rewritten such that it is approximately identical to the \abbrQNIEKF, albeit for an additional correction term.
This enables a richer understanding of the properties of iterated filtering algorithms based on statistical linearization.
\end{abstract}


\IEEEpubid{}
\AddToShipoutPictureBG*{%
	\put(0,20){
		\hspace*{\dimexpr0.075\paperwidth\relax}
		\parbox{.84\paperwidth}{\footnotesize ~\copyright2023 IEEE. Personal use of this material is permitted. Permission from IEEE must be obtained for all other uses, in any current or future media, including reprinting/republishing this material for advertising or promotional purposes, creating new collective works, for resale or redistribution to servers or lists, or reuse of any copyrighted component of this work in other works.}%
}}

\section{Introduction}\label{sec:introduction}
\IEEEPARstart{S}{tate} estimation in discrete--time state--space models with additive Gaussian noise, i.e., models such as
\begin{subequations}
    \begin{align}
        \state_{k+1} &= \dynmod(\state_k) + \pnoise_k\label{eq:dynmodel} & \pnoise_k & \stackrel{\text{\tiny i.i.d.}}{\sim} \Ndist(\vec{0}, \vec{Q})\\
        \obs_k &= \obsmod(\state_k) + \onoise_k\label{eq:obsmodel} & \onoise_k &\stackrel{\text{\tiny i.i.d.}}{\sim} \Ndist(\vec{0},\vec{R})
    \end{align}
\end{subequations}
has been studied extensively for decades.
Here, $\state_k,~\obs_k,~\pnoise_k,~\onoise_k$ are the state, measurement, process noise, and measurement noise, respectively. 
The filtering problem is then to compute the marginal distributions $p(\state_k|\obs_{1:k})$, given a sequence of measurements $\obs_{1:k}$.
In the case $\dynmod$ and $\obsmod$ are linear functions, this is analytically tractable and the solution is given by the Kalman filter, which is the optimal estimator in the mean--squared error sense \cite{kalmanNewApproachLinear1960}.
For nonlinear state--space models, analytical solutions generally do not exist and approximate inference techniques have been developed for these cases.
The extended Kalman filter (\abbrEKF), introduced alongside the original Kalman filter, was a simple way of extending the Kalman filter to the nonlinear case \cite{kalmanNewApproachLinear1960}.
The original \abbrEKF linearizes the dynamical model $\dynmod$ and the measurement model $\obsmod$ at the current state estimate using a first--order Taylor expansion and then applies the standard Kalman filter updates.
Since the \abbrEKF was developed, it has been identified that this choice of linearization point may be suboptimal. This has lead to the development of the iterated extended Kalman filter (\abbrIEKF) \cite{denhamSequentialEstimationWhen1965a, Bell93, Teunissen91}.
This is a family of approximate inference techniques that attempt to find a better linearization point for the measurement model, which boils down to iterating the measurement update a number of times.
This family of inference techniques include the line--search \abbrIEKF, quasi--Newton (\abbrQN) \abbrIEKF (\abbrQNIEKF) etc., which are commonly referred to as damped {\abbrIEKF}s \cite{skoglundExtendedKalmanFilter2015}.

Alongside the development of the \abbrIEKF, other strategies for nonlinear filtering were developed to circumvent the need for analytical linearization.
Particularly, the unscented Kalman filter (\abbrUKF) was developed as a competitive alternative to the \abbrEKF \cite{julierNewApproachFiltering1995, julierGeneralMethodApproximating}.
The \abbrUKF is essentially based on deterministically sampling the prior and propagating a set of \q{sigma points} through the nonlinear function, whereafter an approximate Gaussian distribution can be constructed based on the transformed points.
This is done in both the time update, as well as the measurement update.
The \abbrUKF has since been shown to be equivalent to statistically linearizing the nonlinear models and then applying the standard Kalman recursions \cite{lefebvreCommentNewMethod2002}.
Therefore, both the \abbrUKF and other deterministically sampled sigma--point methods can be interpreted as linearization--based nonlinear filtering algorithms.
Similarly to the \abbrEKF, the \abbrUKF has also been extended to the iterated \abbrUKF (\abbrIUKF) \cite{skoglundIterativeUnscentedKalman2019}.

Recently, another strategy for linearization--based filtering was introduced as yet another alternative to the \abbrEKF and the sigma--point based filters, namely the iterated posterior linearization filter (\abbrIPLF) \cite{garcia-fernandezIteratedStatisticalLinear2014}.
The \abbrIPLF is based on the idea of linearizing around the \emph{posterior} $p(\state_k|\obs_{1:k})$.
As the posterior is not available, the \abbrIPLF constructs an approximate posterior $q_i(\state_k|\obs_{1:k})$ and then iterates the measurement update, each time performing statistical linearization around the current approximate posterior $q_i(\state_k|\obs_{1:k})$.
Similarly to the \abbrIEKF, the \abbrIPLF has been shown to diverge for some particular problems \cite{skoglundIterativeUnscentedKalman2019}, which has led to the development of damped versions of the \abbrIPLF \cite{raitoharjuDampedPosteriorLinearization2018}.

To gain a deeper understanding of the properties of the iterated statistical linearization filters, such as the \abbrIUKF and \abbrIPLF, we seek to connect this family of methods to classical \abbrQN methods, such as the \abbrQNIEKF.
To that end, we firstly find a Hessian correction in the \abbrQNIEKF such that the iterate updates are identical to that of the \abbrIPLF, thereby showing that the \abbrIPLF, \abbrIUKF and other iterated statistical linearization based filters may be viewed as \abbrQN methods.
Secondly, as the necessary Hessian correction has a complicated form, we show that the \abbrIPLF/\abbrIUKF can be rewritten in such a way that an approximate \abbrQN structure appears without the need for a complicated Hessian correction.
However, this secondary correspondence is only approximate as it requires an additional correction term, which nonetheless is fully interpretable.


\section{Preliminaries}\label{sec:preliminaries}
Mathematical preliminaries are restated here for completeness.

\subsection{Statistical Linearization}\label{subsec:statisticallinearization}
Given a nonlinear model
\begin{equation*}
\vec{z} = \vec{g}(\state),
\end{equation*}
we wish to find an affine representation
\begin{equation}
\vec{g}(\state)\approx \vec{A}\state + \vec{b} + \eta,
\end{equation}
with $\eta\sim\Ndist(\eta; \vec{0}, \boldsymbol{\Omega})$.
In this affine representation, there are three free parameters, $\vec{A}, \vec{b}$, and $\boldsymbol{\Omega}$.
Statistical linearization finds these parameters by linearizing w.r.t. a distribution $p(\state)$.
Practically, one may think of this as constructing an affine function that best fits a number of samples of $p(\state)$ transformed through $\vec{g}(\state)$.
Assuming that $p(\state)=\Ndist(\state;\hat{\state},\vec{P})$, statistical linearization selects the affine parameters as
\begin{subequations}\label{eq:statisticallinearization}
\begin{align}
\vec{A} &= \Psi^\top \vec{P}^{-1}, \qquad~\;\vec{b} = \bar{\vec{z}}-\vec{A}\hat{\state}\\
\boldsymbol{\Omega} &= \Phi-\vec{A} \vec{P} \vec{A}^\top, \quad
\bar{\vec{z}} = \mathbb{E}[\vec{g}(\state)]\\
\Psi &= \mathbb{E}[(\state-\hat{\state})(\vec{g}(\state) - \bar{\vec{z}})^\top]\\
\Phi &= \mathbb{E}[(\vec{g}(\state) - \bar{\vec{z}})(\vec{g}(\state) - \bar{\vec{z}})^\top],
\end{align}\noindent
\end{subequations}
where the expectations are taken w.r.t. $p(\state)$.
The major difference from analytical linearization is that ${\boldsymbol{\Omega}\neq 0}$, which implies that the error in the linearization is captured.
Typically, the expectations in \cref{eq:statisticallinearization} are not analytically tractable and thus, practically, one often resorts to some numerical integration technique.

\subsection{Quasi--Newton Optimization}
A general nonlinear least--squares minimization problem is given by
\begin{equation}\label{eq:minproblem}
    \hat{\state} = \argmin_\state V(\state), \quad V(\state) = \frac{1}{2} r(\state)^\top r(\state).
\end{equation}
One particular family of methods for solving these problems, is the Newton family.
This family of methods essentially finds the minimizing argument of \cref{eq:minproblem} by starting at an initial guess $\state_0$ and iterating
\begin{equation}\label{eq:newtoniteration}
    \state_{i+1} = \state_i - (\nabla^2V(\state_i))^{-1}\nabla V(\state_i).
\end{equation}
Here, $\nabla^2V(\state_i)$ and $\nabla V(\state_i)$ are the Hessian and the gradient of $V$ evaluated at $(\state_i)$, respectively. 
Note that convergence of the iterates typically benefit from step--size correction, see \eg, \cite{Nocedal2006Numerical}.
For nonlinear least--squares problems, the gradient and Hessian are given by
\begin{subequations}
    \begin{align}
        \nabla V(\state) &= \vec{J}^\top(\state)r(\state), \quad \vec{J}(\state) = \frac{d r(\vec{s})}{d \vec{s}}\bigg\lvert_{\vec{s}=\state}\\
        \nabla^2 V(\state) &= \vec{J}^\top(\state)\vec{J}(\state) + \sum_{i=1}^{n_r}[r(\state)]_i\nabla^2 [r(\state)]_i
    \end{align}
\end{subequations}
where $[r(\state)]_i$ is the $i$th component of $r(\state)$ and $n_r$ is the dimension of $r(\state)$.
As the Hessian of the cost function can be computationally expensive to evaluate, approximate versions of Newton's method have been developed.
In particular, the Gauss--Newton method approximates the Hessian as 
\begin{equation}\label{eq:gaussnewtonhessian}
    \nabla^2 V(\state) \approx \vec{J}(\state)^\top\vec{J}(\state),
\end{equation}
thus only requiring first--order information. 
As such, it is a Quasi--Newton (\abbrQN) method since it operates as a Newton method with an approximate Hessian.
This approximation may be bad far from the optimum, which may affect convergence.
A remedy is to either approximate the Hessian directly in some other way, or by introducing a correction term to the Gauss--Newton approximate Hessian as
\begin{equation}\label{eq:quasinewtonhessian}
    \nabla^2 V(\state) \approx \vec{J}(\state)^\top\vec{J}(\state) + \vec{T},
\end{equation}
where $\vec{T}$ is supposed to capture second--order information and can be chosen in a variety of ways, see \cite{Nocedal2006Numerical} for an overview. 

\subsection{Quasi--Newton \abbrIEKF}
The cost function for the general \abbrIEKF is of the form \cref{eq:minproblem} with \cite{skoglundExtendedKalmanFilter2015}
\begin{equation}
    r(\state) = \begin{bmatrix}
        \vec{R}^{-1/2}(\obs_k-\obsmod(\state))\\
        \vec{P}_{k|k-1}^{-1/2}(\hat{\state}_{k|k-1}-\state)
    \end{bmatrix}.
\end{equation}
Hence, the Jacobian $\vec{J}(\state)$ is given by
\begin{equation}
    \vec{J}(\state) = -\begin{bmatrix}
        \vec{R}^{-1/2}\vec{H}(\state)\\
        \vec{P}_{k|k-1}^{-1/2}
    \end{bmatrix}, ~ \text{where} ~ \vec{H}(\state) = \frac{d \obsmod(\vec{s})}{d \vec{s}}\bigg\lvert_{\vec{s}=\state}.
\end{equation}
Now, using \cref{eq:newtoniteration} and \cref{eq:quasinewtonhessian}, the \abbrQNIEKF iterate update is given by \cite{skoglundExtendedKalmanFilter2015}
\begin{subequations}\label{eq:qnekf}
    \begin{align}
        \state_{i+1} &= \hat{\state} + \vec{K}^q_i(\obs_k - \obsmod_i - \vec{H}_i\tilde{\state}_i) - \vec{S}^q_i\vec{T}_i\tilde{\state}_i\label{eq:qnupdate}\\
        \vec{P}_{i+1} &= \vec{P} - \vec{P}\vec{H}_i^\top(\vec{H}_i\vec{P}\vec{H}_i^\top+\vec{R})^{-1}\vec{H}_i\vec{P}\label{eq:qncovariance}\\
        \vec{S}^q_i &\triangleq \left( \vec{H}_i^\top\vec{R}_i^{-1}\vec{H}_i + \vec{P}^{-1} + \vec{T}_i \right)^{-1}\\
        \vec{K}^q_i &\triangleq \vec{S}^q_i\vec{H}_i^\top\vec{R}^{-1},
    \end{align}
\end{subequations}
with simplified notation $\tilde{\state}_i=\hat\state - \state_i$, $\vec{h}_i=\obsmod(\state_i)$, $\vec{H}_i=\vec{H}(\state_i)$, and $\vec{P}=\vec{P}_{k|k-1}$.
Further, $\vec{T}_i$ is the current Hessian correction which can be chosen freely, see \cite{skoglundExtendedKalmanFilter2015}.

\subsection{Iterated Posterior Linearization Filter}
The \abbrIPLF was initially developed through minimizing the Kullback--Leibler (\abbrKL) divergence between the true marginal posterior $p(\state_k|\obs_{1:k})$ and an approximation $q(\state_k|\obs_{1:k})$, \ie,
\begin{equation}\label{eq:iplfkl}
    q(\state_k|\obs_{1:k})=\min_q D_{\abbrKL}\left(p \Vert q \right),
\end{equation}
where 
\begin{equation*}
    D_{\abbrKL}\left(p\Vert q\right) = \int p(\state_k|\obs_{1:k}) \log \frac{p(\state_k|\obs_{1:k})}{q(\state_k|\obs_{1:k})}d\state_k.
\end{equation*}
In particular, the divergence is used to find an \q{optimal} affine approximation of the observation model, subsequently used as the \q{surrogate} measurement model.
However, as the objective \cref{eq:iplfkl} is not analytically tractable, and also requires access to the true posterior $p(\state_k|\obs_{1:k})$, the \abbrIPLF approximately minimizes this by starting at some initial approximate posterior $q_0(\state_k|\obs_{1:k})$. The approximate posterior is then iteratively refined until a stopping condition is met.
Essentially, the recursions are similar to an \abbrIEKF using a statistically linearized model.
By adapting the notation of \cite{garcia-fernandezPosteriorLinearizationFilter2015}, the iterate update of the \abbrIPLF can be written as
\begin{subequations}\label{eq:iplf}
    \begin{align}
        \state_{i+1}^{\text{\tiny{IPLF}}} &= \hat{\state} + \vec{K}_i(\obs_k-\obsmod_i-\vec{H}_i\tilde{\state}_i)\label{eq:iplfupdate}\\
        \vec{P}_{i+1} &= \vec{P} - \vec{K}_i\vec{S}_i\vec{K}_i^\top\label{eq:iplfcovariance}\\
        \vec{K}_i &\triangleq \vec{P}_i\vec{H}_i^\top\vec{S}_i\label{eq:iplfkalmangain}\\
        \vec{S}_i &\triangleq \left( \vec{H}_i\vec{P}_i\vec{H}_i^\top + \vec{R} + \Om_i \right)^{-1}\label{eq:iplfinnovationcovariance},
    \end{align}
\end{subequations}
where $\Om_i,~\obsmod_i$ and $\vec{H}_i$ are found through statistical linearization of $\obsmod$.
Note that $\obsmod_i=\vec{b}+\vec{A}\hat{\state}$ and $\vec{H}_i=\vec{A}$, see \cref{subsec:statisticallinearization}.
The iterates are initialized as $\state_i = \hat{\state},~\vec{P}_i=\vec{P}$ and the updates are then iterated until the approximate posterior $q_i(\state_k|\obs_{1:k})$ does not significantly change, as measured by 
\begin{equation*}
    D_{\abbrKL}( q_{i+1} \Vert q_{i} ).
\end{equation*}

Note that by choosing the covariance update \cref{eq:iplfcovariance} as $\vec{P}_{i+1}=\vec{P}$ until the last iteration, the \abbrIUKF presented in \cite{skoglundIterativeUnscentedKalman2019} is obtained. 
We will now consider two relationships between the \abbrIPLF update \cref{eq:iplfupdate} and the \abbrQN--\abbrIEKF update \cref{eq:qnupdate}.

\section{Exact Quasi--Newton}\label{sec:exactqn}
Next, we show that the \abbrIPLF can be viewed as an \emph{exact} \abbrQN method, \ie, that it corresponds to \abbrQN with a particular choice of Hessian correction $\vec{T}_i$.
More precisely, we find a Hessian correction $\vec{T}_i$ such that the \abbrQN--\abbrIEKF update \cref{eq:qnupdate} is equal to the \abbrIPLF update \cref{eq:iplfupdate}.

First, let $\epsilon_i \triangleq \obs_k - \obsmod_i - \vec{H}_i\tilde{\state}_i$. 
Now, setting \cref{eq:qnupdate} equal to \cref{eq:iplfupdate} yields
\begin{align*}
    &\vec{K}^q_i\epsilon_i - \vec{S}^q_i\vec{T}_i\tilde{\state}_i = \vec{K}_i\epsilon_i \iff \\
    &\vec{S}^q_i\vec{H}_i^\top\vec{R}^{-1}\epsilon_i - \vec{S}^q_i\vec{T}_i\tilde{\state}_i = \vec{K}_i\epsilon_i \iff \\
    &\vec{H}_i^\top\vec{R}^{-1}\epsilon_i - \vec{T}_i\tilde{\state}_i = (\vec{S}^q_i)^{-1}\vec{K}_i\epsilon_i \iff \\
    &\vec{H}_i^\top\vec{R}^{-1}\epsilon_i - \vec{T}_i\tilde{\state}_i = ( 
\vec{H}_i^\top\vec{R}^{-1}\vec{H}_i + \vec{P}^{-1} + \vec{T}_i )\vec{K}_i\epsilon_i \iff\\
&\underbrace{(\vec{H}_i^\top\vec{R}^{-1} \!-\! \big(\vec{H}_i^\top\vec{R}^{-1}\vec{H}_i \!+\! \vec{P}^{-1}\big)\vec{K}_i )}_{(*)}\epsilon_i = \vec{T}_i ( \tilde{\state}_i \!+\! \vec{K}_i\epsilon_i ).
\end{align*}
Now, note that $(*)$ can be written as
\begin{align*}
(*) &= (\vec{H}_i^\top\vec{R}^{-1} - \big(\vec{H}_i^\top\vec{R}^{-1}\vec{H}_i + \vec{P}^{-1}\big)\vec{K}_i ) \\
&= (\vec{H}_i^\top\vec{R}^{-1} - \vec{H}_i^\top\vec{R}^{-1}\vec{H}_i\vec{P}_i\vec{H}_i^\top\vec{S}_i - \vec{P}^{-1}\vec{P}_i\vec{H}_i^\top\vec{S}_i ) \\
&= (\vec{H}_i^\top\vec{R}^{-1}\left( \vec{S}_i^{-1} - \vec{H}_i\vec{P}_i\vec{H}_i^\top \right)\vec{S}_i - \vec{P}^{-1}\vec{P}_i\vec{H}_i^\top\vec{S}_i ) \\
&= (\vec{H}_i^\top\vec{R}^{-1}(\vec{R}+\Om_i)\vec{S}_i - \vec{P}^{-1}\vec{P}_i\vec{H}_i^\top\vec{S}_i )\\
&= (\vec{H}_i^\top\vec{S}_i + \vec{H}_i^\top\vec{R}^{-1}\Om_i\vec{S}_i - \vec{P}^{-1}\vec{P}_i\vec{H}_i^\top\vec{S}_i)\\
&= (\vec{H}_i^\top\vec{R}^{-1}\Om_i + \big(\vec{I} - \vec{P}^{-1}\vec{P}_i \big)\vec{H}_i^\top) \vec{S}_i.
\end{align*}
Thus, we have
\begin{equation*}
    \vec{T}_i (\tilde{\state}_i \!+\! \vec{K}_i\epsilon_i) = (\vec{H}_i^\top\vec{R}^{-1}\Om_i + \big(\vec{I} - \vec{P}^{-1}\vec{P}_i \big)\vec{H}_i^\top) \vec{S}_i\epsilon_i.
\end{equation*}
Letting
\begin{subequations}
\begin{align}
    \vec{s}_i &\triangleq \tilde{\state}_i + \vec{K}_i\epsilon_i = \state_{i+1}^{\text{\tiny{IPLF}}} - \state_i\label{eq:direction}\\
    \vec{p}_i &\triangleq (\vec{H}_i^\top\vec{R}^{-1}\Om_i + \big(\vec{I} - \vec{P}^{-1}\vec{P}_i \big)\vec{H}_i^\top) \vec{S}_i\epsilon_i, \label{eq:residualterm}
\end{align}    
\end{subequations}
we have
\begin{equation}\label{eq:secanteq}
    \vec{T}_i\vec{s}_i = \vec{p}_i.
\end{equation}
which is similar to the \emph{secant equation}, see \eg \cite[p. 24]{Nocedal2006Numerical}. Hence, we can follow a similar reasoning and procedure to find a solution.
That is, we impose that $\vec{T}_i$ be symmetric and that it is \q{close} to $\vec{T}_{i-1}$, in some sense.
Thus, we find $\vec{T}_i$ by minimizing
\begin{equation}\label{eq:optproblem}
\begin{alignedat}{2}
    &\min_{\vec{T}} &\quad &\lVert \vec{T}-\vec{T}_{i-1} \rVert_\vec{W}\\
    &\text{subject to} && \vec{T}=\vec{T}^\top, \quad \vec{T}\vec{s}_i=\vec{p}_i.
\end{alignedat}    
\end{equation}
Here, $\lVert \vec{A} \rVert_\vec{W} = \lVert \vec{W}^{1/2}\vec{A}\vec{W}^{1/2} \rVert$ and $\vec{W}$ is a nonsingular symmetric matrix.
Now, let $\vec{s}_i,\vec{p}_i$ be in $\mathbb{R}^n$.
Then, \cite[Theorem 7.3]{Dennis77} states that for any $\vec{c}\in\mathbb{R}^n$ such that $\vec{c}^\top\vec{s}_i>0$ and $\vec{W}\vec{c}=\vec{W}^{-1}\vec{s}_i$, the solution to \cref{eq:optproblem} is given by
\begin{multline}\label{eq:recursion}
        \vec{T}_i = \vec{T}_{i-1} + \frac{(\vec{p}_i-\vec{T}_{i-1}\vec{s}_i)\vec{c}^\top + \vec{c}(\vec{p}_i-\vec{T}_{i-1}\vec{s}_i)^\top}{\vec{c}^\top\vec{s}_i}  \\
        - \frac{(\vec{p}_i-\vec{T}_{i-1}\vec{s}_i)^\top\vec{s}_i}{(\vec{c}^\top\vec{s}_i)^2}\vec{c}\vec{c}^\top.
\end{multline}
In particular, we choose $\vec{c}=\vec{s}_i$ which guarantees $\vec{c}^\top\vec{s}_i=\vec{s}_i^\top\vec{s}_i>0$ as long as $\vec{s}_i\neq 0$.
Note that the case $\vec{s}_i=0$ is not of interest, as this means that $\state_{i+1}=\state_i$ and hence, fixed point convergence has been reached.
The resulting recursion \cref{eq:recursion} is generally known as the Powell symmetric Broyden (\abbrPSB) update \cite[p. 70]{Dennis77}.
However, even though the structure is the same as \abbrPSB, the definition of $\vec{p}_i$ is not. Thus, with these particular definitions, \cref{eq:recursion} does not correspond exactly to \abbrPSB.

Note that we may also follow the typical derivation of the Davidon--Fletcher--Powell (\abbrDFP) update \cite{Nocedal2006Numerical} by imposing that $\vec{T}_i$ be positive definite.
However, this results in the condition $\vec{s}_i^\top\vec{p}_i>0$, which generally is not the case nor is it trivial to guarantee. 

Note that even though the derivation here focuses on the \abbrIPLF, a similar relationship holds for the \abbrICKF, \abbrIUKF, and other iterated sigma--point filters as well.
Regardless, this shows that iterated filtering algorithms based on statistical linearization can be viewed as \abbrQN methods with a particular choice of Hessian correction $\vec{T}_i$.
Further, we can view the captured linearization error $\Om$ as a correction of the Hessian approximation in order to account for lost second order information.
This follows as a direct consequence of our interpretation of the algorithms as \abbrQN methods, which themselves seek to account for lost second order information, through, e.g., \cref{eq:quasinewtonhessian}.
Next, we interpret the Hessian correction \cref{eq:recursion} by considering the special case of the \abbrIUKF and \abbrICKF, which simplifies \cref{eq:residualterm}.

\subsection{\abbrIUKF/\abbrICKF interpretation}\label{subsec:iukfinterpretation}
The \abbrIUKF \cite{skoglundIterativeUnscentedKalman2019} essentially performs the updates \cref{eq:iplfupdate}, but keeps $\vec{P}_{i+1}=\vec{P}$ fixed until the last iteration.
In this case, \cref{eq:residualterm} simplifies to
\begin{equation}\label{eq:iukfresidualterm}
      \vec{p}_i = \vec{H}_i^\top\vec{R}^{-1}\Om_i\vec{S}_i\epsilon_i.
\end{equation}
Henceforth, we assume that $\vec{s}_i$ is constant and given, such that we can interpret the correction \cref{eq:recursion} only in terms of changes in \cref{eq:iukfresidualterm}.
Inspecting the components of \cref{eq:iukfresidualterm}, it is \q{weighted} with $\vec{R}^{-1}\Om_i\vec{S}_i$. 
Hence, with decreasing measurement uncertainty $\vec{R}$, the Hessian correction grows \q{larger}, as the measurement model is more precise.
Similarly, as the linearization error $\Om_i$ grows, the Hessian correction also increases, which makes sense as it indicates that the measurement function $\obsmod$ is highly nonlinear and the Hessian approximation \cref{eq:gaussnewtonhessian} is most likely poor and needs more correction.
Also, note that if $\Om_i=\vec{0}$, \ie, if the model is completely linear at the current iterate, the Hessian is only corrected according to the iterate difference $\vec{s}_i$.
Lastly, as the innovation covariance $\vec{S}_i^{-1}$ decreases, the correction grows, essentially also indicating that the measurement carries a lot of information that can be exploited.
This interpretation should approximately hold for the \abbrIPLF as well, as usually $\vec{P}^{-1}\vec{P}_i\approx \vec{I}$.
However, a detailed analysis of the exact behavior of the \abbrIPLF is non--trivial, as $\vec{H}_i, \Om_i$ and $\vec{S}_i$ all depend on the previous iterate $\vec{P}_{i}$.

\section{Approximate Quasi--Newton}\label{sec:approxqn}
As an alternative to the exact view in \cref{sec:exactqn}, we can also view the \abbrIPLF as an \emph{approximate} \abbrQN method.
Essentially, it boils down to modifying \cref{eq:iplfupdate} such that it approximately takes the form \cref{eq:qnupdate}.

Start with \cref{eq:iplfinnovationcovariance} and write
\begin{multline*}
    \vec{S}_i = \left(\vec{H}_i\vec{P}_i\vec{H}_i^\top + \vec{R} + \Om_i\right)^{-1} = (\vec{R} + \Om_i)^{-1} \big(\vec{I} - \\\vec{H}_i(\vec{H}_i^\top(\vec{R}+\Om_i)^{-1}\vec{H}_i + \vec{P}_i^{-1})^{-1}\vec{H}_i^\top(\vec{R} + \Om_i)^{-1}\big).
\end{multline*}
Now, note that 
\begin{multline*}
    (\vec{H}_i^\top(\vec{R} + \Om_i)^{-1}\vec{H}_i + \vec{P}_i^{-1})^{-1} \\
    = (\vec{H}_i^\top\vec{R}^{-1}\vec{H}_i + \vec{P}_i^{-1} - \vec{H}_i^\top\vec{R}^{-1}(\vec{R}^{-1} + \Om_i^{-1})^{-1}\vec{R}^{-1}\vec{H}_i)^{-1} \\
    = (\vec{H}_i^\top\vec{R}^{-1}\vec{H}_i + \vec{P}_i^{-1} +\tilde{\vec{T}}_i )^{-1} \triangleq \tilde{\vec{S}}^q_i,
\end{multline*}
with 
\begin{align*}
    \tilde{\vec{T}}_i &\triangleq - \vec{H}_i^\top\vec{R}^{-1}(\vec{R}^{-1} + \Om_i^{-1})^{-1}\vec{R}^{-1}\vec{H}_i \\
    &= -\vec{H}_i^\top \left( \vec{R}^{-1} - \left( \vec{R} + \Om_i \right)^{-1} \right)\vec{H}_i.
\end{align*}
Plugging this into \cref{eq:iplfkalmangain} yields
\begin{multline*}
    \vec{P}_i\vec{H}_i^\top\left( \vec{I} - (\vec{R} + \Om_i)^{-1}\vec{H}_i\tilde{\vec{S}}^q_i\vec{H}_i^\top \right)(\vec{R} + \Om_i)^{-1} \\
    = \vec{P}_i\left( (\tilde{\vec{S}}^q_i)^{-1} - \vec{H}_i^\top(\vec{R} + \Om_i)^{-1}\vec{H}_i \right)\tilde{\vec{S}}^q_i\vec{H}_i^\top(\vec{R} + \Om_i)^{-1} \\
    = \tilde{\vec{S}}^q_i\vec{H}_i^\top(\vec{R} + \Om_i)^{-1} \\
    = \underbrace{\tilde{\vec{S}}^q_i\vec{H}_i^\top \vec{R}^{-1}}_{\triangleq \tilde{\vec{K}}^q_i} - \tilde{\vec{S}}^q_i\vec{H}_i^\top \vec{R}^{-1}(\vec{R}^{-1} + \Om_i^{-1})^{-1}\vec{R}^{-1}.
\end{multline*}
Hence, with $\epsilon_i=\obs_k - \obsmod_i - \vec{H}_i\tilde{\state}_i$, the iterate update becomes
\begin{multline*}
    \state_{i+1} = \hat{\state} + \tilde{\vec{K}}^q_i\epsilon_i
    - \tilde{\vec{S}}^q_i\vec{H}_i^\top \vec{R}^{-1}(\vec{R}^{-1}+\Om_i^{-1})^{-1}\vec{R}^{-1}\epsilon_i \\
    = \hat{\state} + \tilde{\vec{K}}^q_i\epsilon_i
    - \tilde{\vec{S}}^q_i
    \overbrace{(-\vec{H}_i^\top \vec{R}^{-1}(\vec{R}^{-1}+\Om_i^{-1})^{-1}\vec{R}^{-1}\vec{H}_i)}^{\tilde{\vec{T}}_i}\tilde{\state}_i \\
    - \tilde{\vec{S}}^q_i\vec{H}_i^\top \vec{R}^{-1}(\vec{R}^{-1} + \Om_i^{-1})^{-1}\vec{R}^{-1}(\obs_k-\obsmod_i) \\
    = \hat{\state} 
    +\tilde{\vec{K}}^q_i\epsilon_i
    - \tilde{\vec{S}}^q_i\tilde{\vec{T}}_i\tilde{\state}_i 
    - \underbrace{\tilde{\vec{S}}^q_i\vec{H}_i^\top (\vec{R}^{-1} - (\vec{R}+\Om_i)^{-1})(\obs_k-\obsmod_i)}_{\triangleq \Delta_i}.
\end{multline*}
Hence, the \abbrIPLF can be interpreted as performing modified Quasi-Newton with a specific choice of Hessian correction $\vec{T}_i$ and an additional correction term in the iterate update.
The additional term $\Delta_i$ can be viewed as a correction of the iterate based on a $0$th order Taylor expansion of the measurement model at the current iterate $\state_i$.
The step is further weighted by $\vec{R}^{-1}-(\vec{R}+\Om_i)^{-1}$, which can be interpreted as a measure of how close to linear the model is. 
Particularly, with $\Om_i=\vec{0}$, the model is completely linear \emph{at the current iterate} $\state_i$, and $\Delta_i$ and $\tilde{\vec{T}}_i$ thus collapse to $\vec{0}$.
This is, of course, completely natural, as the Hessian of a linear model is identically $\vec{0}$.
This means that there is no additional information to extract from the curvature of the measurement model at, and around, the current iterate $\state_i$.
Further, the iterate update collapses to that of the standard Kalman filter, a desirable property of nonlinear filters applied to linear models.

On the other hand, if the model is highly nonlinear, such that $\Om_i$ is much larger than $\vec{R}$, the weighting becomes ${\vec{R}^{-1}-(\vec{R}+\Om_i)^{-1}\approx \vec{R}^{-1}}$ which yields
\begin{equation*}
    \tilde{\vec{T}}_i = -\vec{H}_i^\top\vec{R}^{-1}\vec{H}_i, \quad 
    \Delta_i = \tilde{\vec{S}}^q_i\vec{H}_i^\top\vec{R}^{-1}(\obs_k-\obsmod_i).
\end{equation*}
The terms in the iterate update related to $\tilde{\vec{T}}_i$ and $\Delta_i$ become
\begin{multline*}
    - \tilde{\vec{S}}^q_i(-\vec{H}_i^\top\vec{R}^{-1}\vec{H}_i)\tilde{\state}_i 
    - \tilde{\vec{S}}^q_i\vec{H}_i^\top\vec{R}^{-1}(\obs_k-\obsmod_i)\\
    = - \tilde{\vec{S}}^q_i\vec{H}_i^\top\vec{R}^{-1}(\obs_k-\obsmod_i-\vec{H}_i\tilde{\state}_i)=-\tilde{\vec{K}}_i^q\epsilon_i.
\end{multline*}
Hence, the iterate update collapses to $\state_{i+1}=\hat{\state}$ which means that when the model is highly nonlinear, the \abbrIPLF (and the \abbrIUKF/\abbrICKF) will essentially avoid updating the iterate. 
This is reasonable, as it means that an approximate linear (affine) model is not an appropriate choice.
This also means that iterated filters based on statistical linearization are automatically ``cautious'' in highly nonlinear regions of the measurement model, a feature not present in the standard \abbrIEKF for instance.
In particular, away from the limiting cases, these algorithms still adapt the step length depending on how well the approximate linear (affine) model approximates the true model.
Further, this means that there is already a built--in Hessian correction in the statistically linearized filters and any \abbrQN versions thereof should take this into account when designing their respective Hessian approximations.

\section{Conclusion}\label{sec:introduction}
In this letter, we have shown that iterated filtering algorithms based on statistical linearization, such as the \abbrIUKF and the \abbrIPLF, can be interpreted as \abbrQN methods with a particular choice of Hessian correction.
Through this connection, we hope to enable a richer understanding of the properties of the \abbrIUKF, \abbrIPLF, and other iterated sigma--point filters.
In particular, it allows us to view the use of statistical linearization in iterated filters as an adaptation of the Hessian approximation to account for lost second order information, a direct consequence of the interpretation as \abbrQN methods.
This interpretation can and should be exploited when designing damped versions of iterated statistically linearized filters.
The correspondence may also prove useful in determining the convergence differences between the \abbrIUKF and \abbrIPLF as they differ only slightly from one another.

\bibliographystyle{IEEEtran}
\bibliography{ms}
\end{document}